\documentclass[aps, prb,superscriptaddress,
twocolumn,showpacs,preprintnumbers,amsmath,amssymb]{revtex4}

\usepackage{color}
\usepackage{amsmath}
\usepackage{amssymb}
\usepackage{epsfig}
\usepackage{graphicx}
\usepackage{wasysym}
\usepackage{bbm}

\newcommand \be{\begin{equation}}
\newcommand \ee{\end{equation}}
\newcommand \bes{\begin{equation*}} 
\newcommand \ees{\end{equation*}}
\newcommand \bea{\begin{eqnarray}}
\newcommand \eea{\end{eqnarray}}
\newcommand \beas{\begin{eqnarray*}} 
\newcommand \eeas{\end{eqnarray*}}
\newcommand \bfg{\begin{figure}}
\newcommand \efg{\end{figure}}
\newcommand \bfgs{\begin{figure*}} 
\newcommand \efgs{\end{figure*}}
\newcommand \bwt{\begin{widetext}}
\newcommand \ewt{\end{widetext}}

\newcommand \fig[1]{Fig.~\ref{#1}}

\newcommand \veck{\vec{k}}

\newcommand \idmat{{\bf 1}}

\newcommand \citejournal[4]{#1, {\bf #2}, #3 (#4)}

\begin{document}
\title{Nodal Spin Density Wave and band topology of the FeAs based
materials
}
\author{Ying Ran}
\author{Fa Wang}
\author{Hui Zhai}
\author{Ashvin Vishwanath}
\author{Dung-Hai Lee}
\affiliation{Department of Physics, University of California at
Berkeley, Berkeley, CA 94720, USA} \affiliation{Materials Sciences
Division, Lawrence Berkeley National Laboratory, Berkeley, CA 94720,
USA}

\date{\today}

\begin{abstract}
  The recently discovered FeAs-based materials exhibit a $(\pi,0)$ Spin Density
Wave (SDW) in the undoped state, which gives way to
superconductivity upon doping. Here we show that due to an
interesting topological feature of the band structure, the SDW state
cannot acquire a full gap. This is demonstrated within the SDW
mean-field theory of both a simplified two band model and a more
realistic 5-band model. The positions of the nodes are different in
the two models and can be used to detected the validity of each
model.
\end{abstract}

\maketitle

\section{Introduction}

Since the discovery of superconductivity in La$_{1-x}$F$_x$FeAs at
$T_c=26K$\cite{Kamihara}, there has been mounting excitement
associated both with the rapidly increasing $T_c$ when La is
substituted by other lanthanoids \cite{Tcincrease}, as well as the
similarity with the cuprate superconductors. As with the cuprates,
the FeAs materials are quasi 2D square lattice based transition
metal compounds, which are magnetically ordered at stoichiometry. On
doping, in both cases, the magnetism is replaced by
superconductivity.

However, there are several significant differences. Most
importantly, the FeAs stoichiometric compounds are not
insulating\cite{Kamihara,transport}. Also, the magnetic order here
is along the $(\pi,0)$ direction, and has a small moment
\cite{Neutrons}. In contrast to the cuprates, where only the
d$_{x^2-y^2}$ orbital of the five Cu d-orbitals is important, the
multi-orbital nature of the FeAs materials have been emphasized in
several recent calculations \cite{LDA, multiband,SiQ}. In this paper
we point out a topological aspect of the band structure, closely
connected with the multi-orbital nature of the material, that has
important ramifications for the phases in this system. In particular
we show that a symmetry enforced band degeneracy at high symmetry
points in the Brillouin zone leads to a band structure with
nontrivial topology. This can be quantified in terms of a
`vorticity' quantum number.
An example of vorticity $\pm 1$ is the Dirac node. Here the
vorticity takes on values $\pm 2$.
While such degenerate points occur in the band structure of other
multi-orbital systems, here they actually occur close to the Fermi
level.
In this article we discuss both a simplified 2-band model and a
realistic 5-band model where this is explicitly realized.

An important consequence of this nontrivial band topology is that it
leads to an unusual Spin Density Wave (SDW) state that is
necessarily gapless. Specifically, the SDW wavevector connects hole
pockets with vorticity $=\pm 2$ with electron pockets with zero
vorticity. This mismatch forces nodes in the SDW gap function even
in the presence of perfect nesting. Away from nesting the nodes are
offset from the fermi energy, resulting in fermi pockets. We term
such magnetic order `nodal-SDW'. We emphasize that the topological
feature required for the nodal SDW state exists both in other
simplified models in the literature\cite{ScalapinoZhang,LeeWen} as
well as first-principles band structures \cite{LDA}.

In the following we first derive  and study a simplified two-orbital
tight-binding model motivated by the quantum chemistry. Other
studies \cite{ScalapinoZhang,Li,SiQ} have also focused on two
orbital models. However it has been argued\cite{LeeWen,LDA} that one
needs at least three orbitals to accurately reproduce the LDA band
structure and Fermi surfaces, e.g. location of the hole pockets in
the Brillouin Zone. We then directly study the more general and
realistic 5-band model based on LDA calculation\cite{LDA}. We
consider the stoichiometric compound (zero doping), and study the
mean-field SDW phase. This mean field analysis confirms the
existence of nodes in the SDW gap function in both models
.  We also established the topological stability of the nodes. The
location of the nodes in momentum space, and the associated fermi
surface topologies, however, are different in the two models. This
can be used to detect the validity of 2-band or 5-band model. A
recent numerical renormalization group  study by us has found
precisely this nodal structure in the two-band model. The general
requirements for the existence of these nodes are also discussed.

\section{Nodal SDW in the 2-band model}
\subsection{2-band microscopic model and band structure}
\begin{figure}
\includegraphics[width=0.28\textwidth]{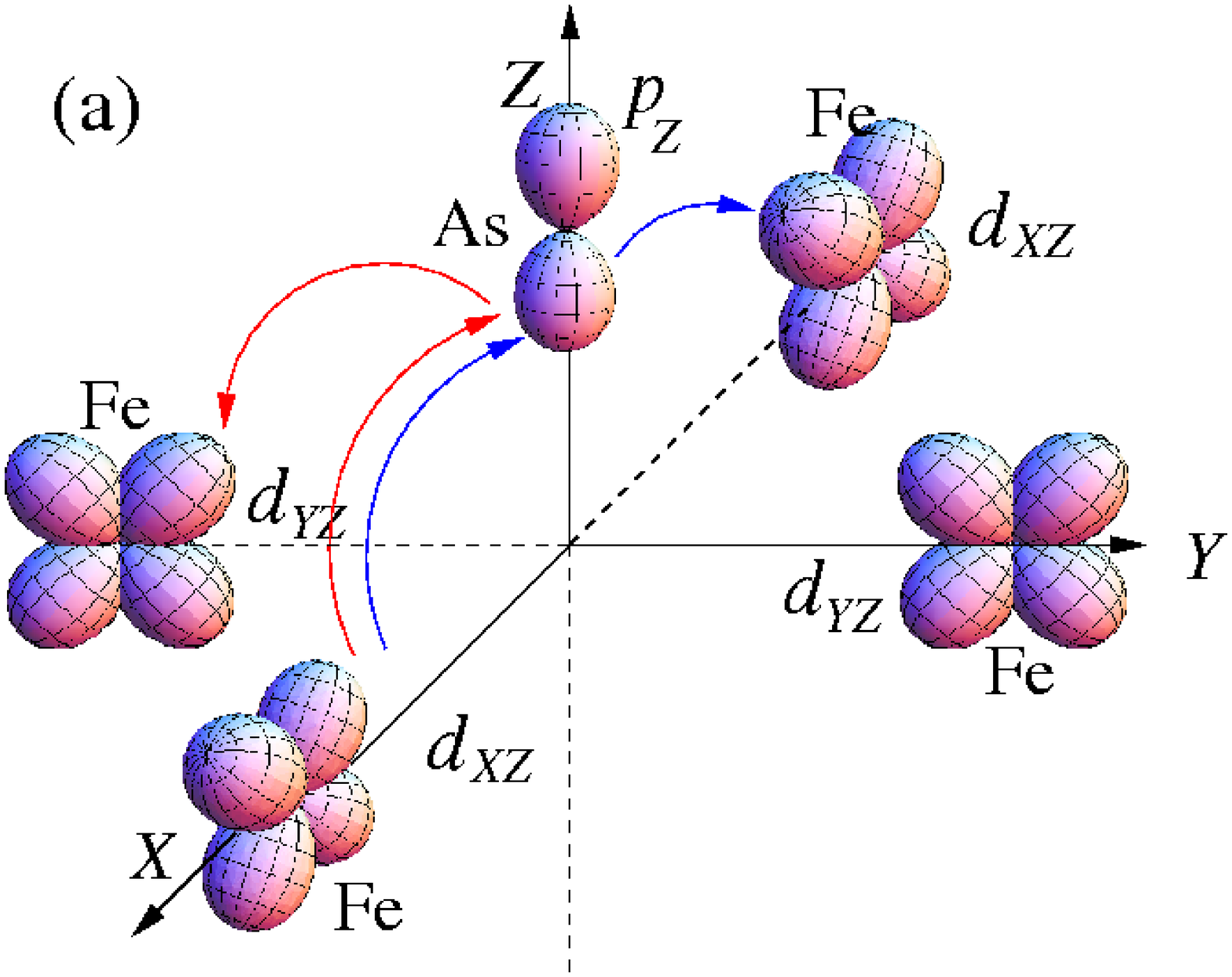}\;\includegraphics[width=0.2\textwidth]{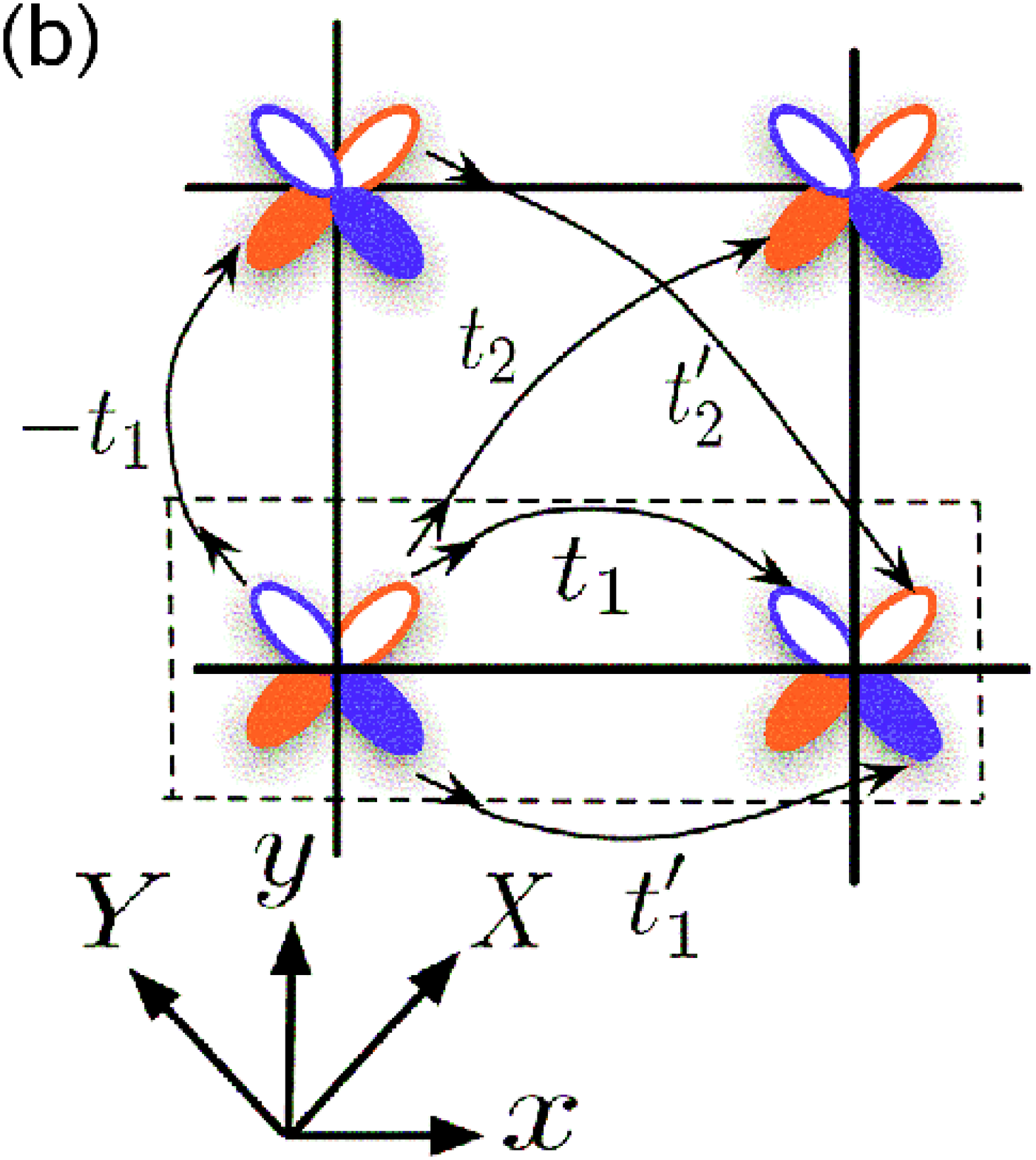}
\caption{(color online) (a) The $3d_{XZ}$ and $3d_{YZ}$ orbitals of the Fe atoms
and the $4p_Z$ orbital of the As atom. The blue(red) arrows
represents one second order process contributing to $t_2$($t_1$).
(b) The tight-binding hoppings between the d-orbitals.}
\label{fig:hoppings}
\end{figure}

The Fe atoms form a square lattice whose principal axes are denoted
as $x$ and $y$, and the crystal structure axes are labeled as $X$
and $Y$[\fig{fig:hoppings}(b)].
Let us first assume that Fe $3d_{xz}$ and $3d_{yz}$ are the relevant
orbitals to describe the low energy physics of this material.
Because of the tetragonal symmetry they are locally degenerate, and
we use their linear combinations $3d_{XZ}$ and $3d_{YZ}$ as our
basis since they have clear symmetry when hybridized with the
nearest As 4p orbitals. Other $d$ orbitals will be ignored here, but
included in a later section.

In a simple chemistry picture[\fig{fig:hoppings}(a)] all symmetry
allowed Fe 3d-As 4p hybridizations are assumed to dominate over
direct Fe 3d hybridization. This naturally leads to large
nearest-neighbor(NN) orbital-changing hopping $t_1$ (see
\fig{fig:hoppings}(b), note that $t_1$ has opposite signs between
vertical and horizontal bonds), next-nearest-neighbor
orbital-preserving hoppings $t_2$ and $t_2'$. If we further assume
that the hybridizations between  $3d_{XZ,YZ}$ and $4p_{Z}$, shown in
\fig{fig:hoppings}(a) dominate, we expect $t_2 \sim t_1
> |t_2'|$, while the direct hopping $ t_1'$ is expected to be much smaller than
these three \footnote{Ref. \cite{Li} similar reasoning is employed
but $3d_{XZ,YZ} \,-\, 4p_Z$ hybridization is neglected which we
believe is dominant. The parameters of Ref\cite{ScalapinoZhang} are
related to ours via
$t_1=(t_2-t_1)/2;\,t_1'=-(t_1+t_2)/2;\,t_2=t_4-t_3;\,t_2'=-t_4-t_3$}.
Given the empirically observed SDW order, we take $t_2-t_2'>|t_1|$,
which leads to nested electron and hole pockets at $(\pi,0)$.

To simplify the notation we will use $d_1$ and $d_2$ for electron
operators associated with the XZ and YZ orbitals. The tight binding
Hamiltonian has {\em only} one Fe atom per unit cell, and we choose
the Brillouin zone accordingly. Many other studies work with a two
Fe atom unit cell, a comparison requires an appropriate folding of
the Brillouin zone of the present study.  Its Fourier transform is
(independent of spin) \be H_0= \sum_{\veck}
\begin{pmatrix}d^\dagger_{1,\veck} & d^\dagger_{2,\veck}\end{pmatrix} K(\veck)
\begin{pmatrix} d_{1,\veck} \\ d_{2,\veck}\end{pmatrix}
\ee where the sum is over
$k_x\in[-\pi,\pi),\ k_y\in[-\pi,\pi)$, the $2\times 2$ matrix
$K(\veck)$ is
\begin{align}
&K(k_x,k_y)= 2 t_1(\cos k_x-\cos k_y)\tau_1 - 2(t_2-t_2')\sin k_x \sin k_y\tau_3 \notag\\
& +[ 2(t_2+t_2')\cos k_x\cos k_y+2 t_1'(\cos k_x+\cos
k_y)]\cdot\idmat\label{eq:2-band}
\end{align}
and $\tau_{1,2,3}$ are Pauli matrices. As argued previously we
expect $ t_2
>t_1\gg t_2',t_1'>0$. The two energy eigenvalues of
Eq.(\ref{eq:2-band}) are:
\begin{align}
&E_{\pm}(\vec k)=  2(t_2+t_2')\cos k_x \cos k_y+2t_1'(\cos k_x+\cos k_y)\notag\\
&\pm 2\sqrt{ t_1^2[\cos k_x-\cos k_y]^2 + (t_2-t_2')^2\sin^2
k_x\sin^2 k_y}\label{eq:dispersion}
\end{align}

\begin{figure}
 \includegraphics[width=0.2\textwidth]{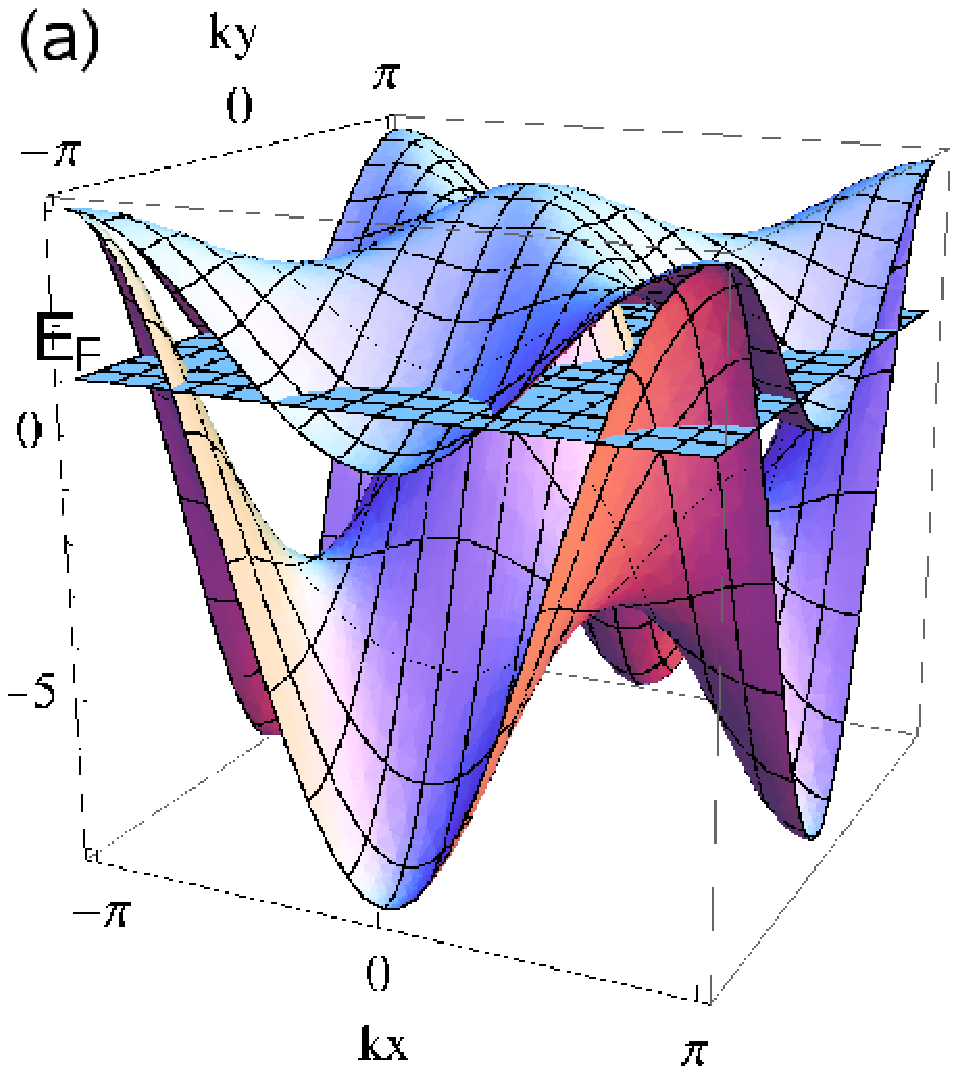}\;\;\;\;\;\;\;\includegraphics[width=0.25\textwidth]{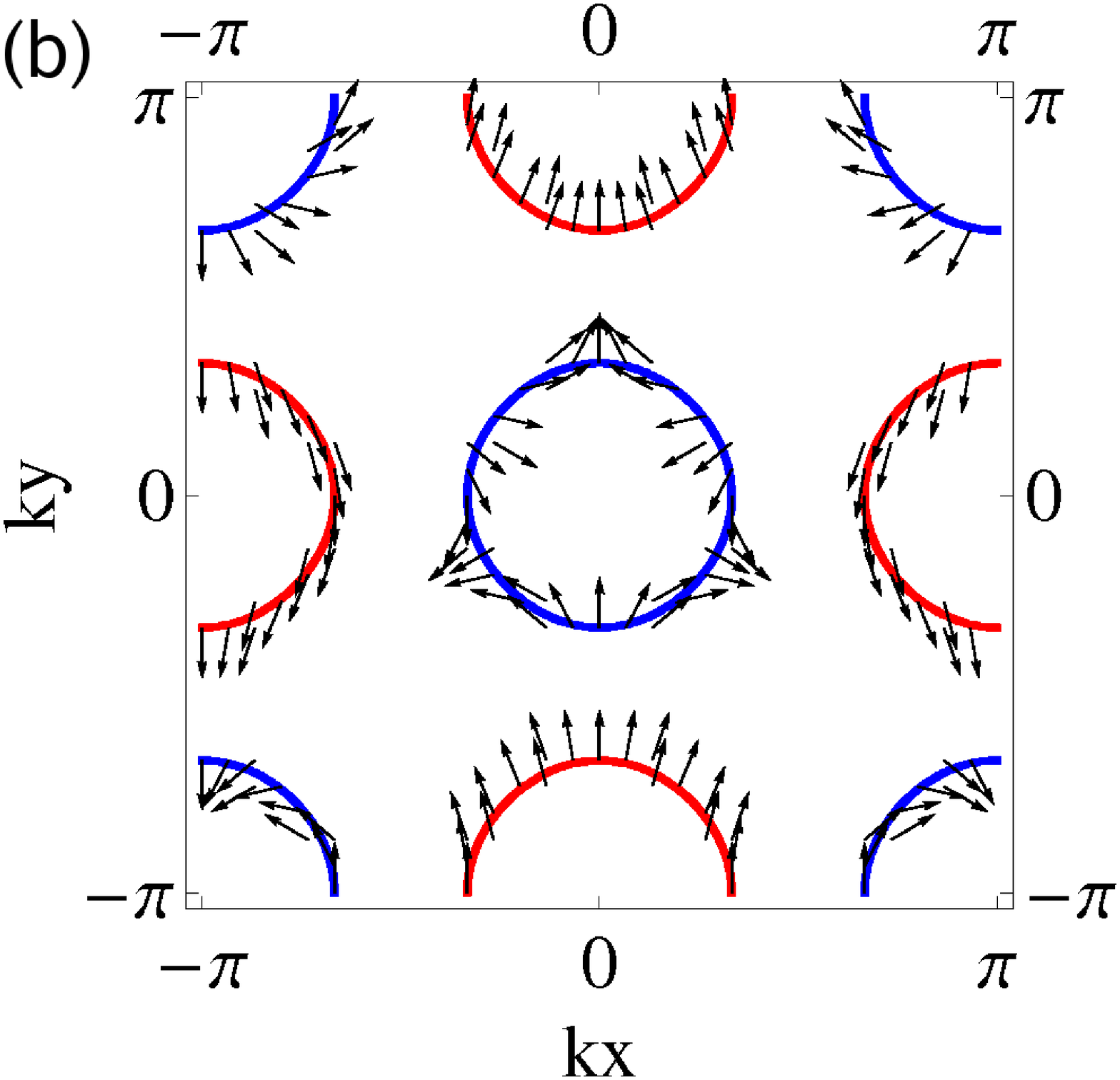}
\caption{(color online) For $t_1=1,t_2=1.7,t_2'=0.3,t_1'=0$  we plot (a): the
dispersion $E_{\pm}(\vec k)$ as shown in Eq.(\ref{eq:dispersion})
together with the half-filling fermi energy $E_F$ , and (b): the
perfectly nested fermi surfaces as given by Eq.(\ref{eq:pocket}) and
the direction of vector $(\cos\phi_{\vec k},\sin\phi_{\vec k})$
defined by rewriting $K(k_x,k_y)$ in Eq.(\ref{eq:2-band}) as $K(k_x,k_y)=a(\vec k)\idmat+b(\vec k)\big[\sin\phi_{\vec
k}\tau_1+\cos\phi_{\vec k}\tau_3\big]$ where $b(\vec k)>0$.} \label{fig:perfect-nesting}
\end{figure}

In Fig.\ref{fig:perfect-nesting}(a) we plot the band structure
Eq.(\ref{eq:dispersion}). At half-filling the fermi level will cut
out two hole-like fermi surfaces around $(0,0)$ and $(\pi,\pi)$, and
two electron-like fermi surfaces around $(\pi,0)$ and $(0,\pi)$ (see
Fig.\ref{fig:perfect-nesting}(b)). Note the band touchings at
$(0,0)$ and $(\pi,\pi)$. This endows the hole Fermi surfaces with
'vorticity' $\pm 2$, where the spinor describing the admixture of
$3d_{XZ}$ and $3d_{YZ}$ orbitals rotates twice on encircling these
Fermi surfaces. The simultaneous presence of inversion and time
reversal symmetry in this band structure allows us to choose at each
$k$ point, real spinor wavefunctions which are hence confined to a
plane. Vorticity in this spinor field is therefore topologically
protected - the singularity at the vortex center forces the orbital
degeneracy at $(0,0)$ and $(\pi,\pi)$. In contrast, the electron
Fermi surfaces are topologically trivial, with no winding  as
shown in Figure \ref{fig:perfect-nesting}b. This topological
characterization of the fermi pockets is also present in more
realistic LDA calculations \cite{LDA, multiband,SiQ}, although it
has not been previously commented upon.

At half filling, the total electron pocket area equals the total
hole pocket area. If the pockets are every small, i.e. $t_2\approx
t_1$, Taylor expansion of Eq.(\ref{eq:dispersion}) gives four nearly
circular fermi pockets with same area. Therefore the electron and
hole pockets are nested at momentum $(\pi,0)$/$(0,\pi)$.

It is interesting to consider for a moment the case $t_1'=0$. Then,
the model decouples into two independent $t_1-t_2-t_2'$ checkerboard
models. Each give rise to one hole and one electron pocket,
separated by $(\pi,\pi)$ from the other electron-hole pocket pair.
Unexpectedly, the electron and hole pockets are {\em precisely
nested} with momentum $(\pi,0)$/$(0,\pi)$, as long as $t_1'=0$, even
when the pockets are large and non-circular.
One can check that at the half filling Fermi energy
$E_F=\frac{2t_1^2}{t_2+t_2'}$, the fermi surface wavevectors $\vec
k^F$ of the two bands: $E_{-}(\vec k^F)=E_F$ and $E_{+}(\vec
k^F+(\pi,0))=E_F$ satisfy exactly the same condition:
\begin{align}
&\big[ (t_2+t_2')^2 \cos^2 k_x^F -t_1^2\big] \big[ (t_2+t_2')^2 \cos^2 k^F_y -t_1^2 \big] \notag\\
=& (t_2+t_2')^2(t_2-t_2')^2 \sin^2 k^F_x \sin^2 k^F_y
.\label{eq:pocket}
\end{align}
When $t_1'>0$ the two hole pockets have different sizes. The
electron pocket around $(\pi,0)$ ($(0,\pi)$) is elongated along the
$k_y$ ($k_x$) direction and the perfect nesting is lost.

\subsection{Mean Field Study of the SDW order in the 2-band model} Now we include on-site
interactions in an extended two-band Hubbard model with $H=H_0+H_I$
where, as in Ref.\onlinecite{LeeWen}:
\begin{align}
 H_{I}&=\frac{U}{2}\sum_{i}(n_{i1}^2+n_{i2}^2)+(U-2J)\sum_i n_{i1}n_{i2}\notag\\
&+J\sum_{i}^{\alpha,\beta=\uparrow,\downarrow} d_{i1,\alpha}^{\dagger}d_{i2,\beta}^{\dagger}d_{i1,\beta}d_{i2,\alpha}\notag\\
&+J\sum_{i}\big(d_{i1,\uparrow}^{\dagger}d_{i1,\downarrow}^{\dagger}d_{i2,\downarrow}d_{i2,\uparrow}+h.c.\big)
\end{align}
where the first and second terms are the intra-orbital and
inter-orbital Coulomb repulsions. The third term the Hund's coupling
and the fourth term is the inter-orbital pair hopping. First
principles calculations \cite{LDA} suggest that the FeAs material is
in the intermediate coupling regime and we choose $t=1$, $t_1'=0.2$,
$t_2=1.7$, $t_2'=0.3$ for our Hopping Hamiltonian $H_0$, where we
estimate $t_1 \sim 0.3eV$ to get the right bandwidth. Also, we use
$U=4$ ($1.2eV$), $J=0.4$, a factor of three smaller than in \cite{LDA}
to get reasonable SDW transition temperatures and moments compatible
with experiments.

For small $t_1'$, where good nesting prevails and in the presence of
repulsive onsite interactions, it is natural to consider SDW order
at wave-vector $(\pi,0)$. Then,
$H_{SDW}=M_{ab}\sum_i(-)^{i_x}(d_{i,a\uparrow}^{\dagger}d_{i,b\uparrow}-d_{i,a\downarrow}^{\dagger}d_{i,b\downarrow})$
where the spin direction is assumed to be along the $S_z$ axis.
Because of the multi orbital nature of the system, different flavors
of SDW are allowed, described by the Hermitian matrix $M_{ab}$,
which may be parameterized by four real numbers
$M_{ab}=[\phi_0\tau_0+\phi_1\tau_1+\phi_2\tau_2+\phi_3\tau_3]_{ab}$
where $(\tau_0)_{ab}=\delta_{ab}$.
 We perform the finite temperature mean-field study by using a trial
density matrix of the mean-field Hamiltonian $H_{MF}=H_0+H_{SDW}$.
We construct a trial free energy based on this mean-field density
matrix: $F_{trial}(\phi_0,\phi_1,\phi_2,\phi_3)=F_{MF}+\langle
H_{I}-H_{SDW}\rangle_{MF}$ where $F_{MF}$ is the free energy of the
free fermion system described by $H_{MF}$. The Feynman
inequality\cite{Feynman} $F \leq F_{trial}$, implies that we need to
minimize $F_{trial}$ over the mean field parameters $\phi_i$ keeping
the electron density fixed.

\begin{figure}
\begin{center}
\includegraphics[width=0.45\textwidth]{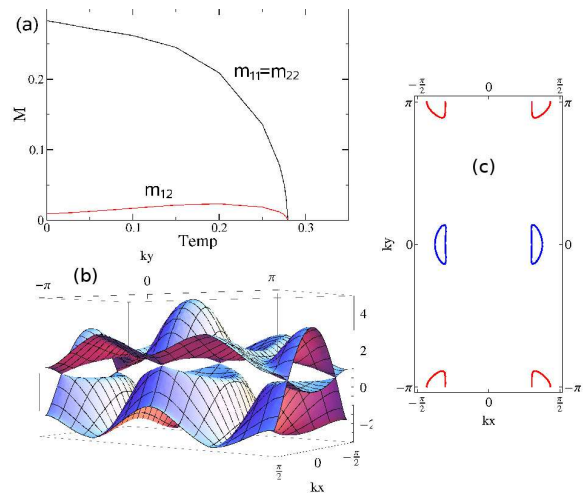}
\end{center}
\caption{(color online) For $t_1=1$, $t_1'=0.2$, $t_2=1.7$, $t_2'=0.3$, $U=4$ and
$J=0.4$ and half-filling, we show (a): the magnetization
$\langle\hat m_{11}\rangle=\langle\hat m_{22}\rangle$ and
$\langle\hat m_{12}\rangle$  of the SDW phase as a function of
temperature. The mean-field calculation is done on 20 by 20
lattice.(b): the zero temperature SDW band structure. We plot the
dispersion of two lowest energy bands in the reduced Brillouin zone.
(c): the fermi pockets. The four nodes are not at the same energy
and therefore the half-filled system has small fermi pockets. The
two nodes close to $k_y=0$ are hole doped (blue pockets) and the two
nodes close to $k_y=\pi$ are electron doped (red pockets).  } \label{fig:sdw}
\end{figure}

Implementing this we find that the model $H_0+H_I$ has a unique SDW
phase at low temperature characterized by $\phi_0\neq0,\phi_1\neq0$. The symmetry of this phase is consistent with the regular $(\pi,0)$ magnetic order depicted in Fig.\ref{fig:sdw-pattern}.
If we denote the order parameter operators $\hat
m_{ab}=d_{i,a\uparrow}^{\dagger}d_{i,b\uparrow}-d_{i,a\downarrow}^{\dagger}d_{i,b\downarrow}$,
in Fig.\ref{fig:sdw}(a) we plot these magnetizations as a function
of temperature and it is clear that $\langle \hat
m_{11}\rangle=\langle\hat m_{22}\rangle\gg\langle\hat
m_{12}\rangle$.  The fact that the $\phi_2$ and $\phi_3$ orders are
not mixed in can be understood from symmetry. Under reflections
$P_y$ about the y-axis crossing the plaquette center, $\phi_0$ and
$\phi_1$ transform differently from $\phi_3$. Under time reversal,
$\phi_2$ order transforms differently from a SDW, and in fact
describes a spin-orbital locked state.

The band structure resulting from this SDW is shown in
Fig.\ref{fig:sdw}(d), where electron and hole pockets in the same
direction as the SDW axis are visible for each pair of Fermi
surfaces. Even with perfect nesting, the Fermi surfaces are {\em
not} fully gapped, instead there are Dirac nodes as in
Fig.\ref{fig:sdw}(b).  In the absence of nesting these Dirac nodes
are at slightly different energies and the half-filled system has
small fermi pockets. As shown below, the presence of such pockets
are required under fairly general conditions
. \emph{This is the main point of the current study.} The presence
of such pockets are detectable by ARPES, and also, due to the
velocity anisotropy present for the nodes, via conductivity
measurements. Based on Drude's formula we compute the DC
conductivities in $x$ and $y$ directions for a mean-field SDW state
with nodes along the $x$ direction and with a $T=0$ magnetic moment
of $0.3\mu_B$ per Fe atom \cite{Neutrons}. We find
$\sigma_{xx}/\sigma_{yy}=6.2$. Any state with broken rotation
symmetry would have conductivity anisotropy, but the large value
here stems from the nodal structure discussed.

For the other SDW type orders,  $\phi_2$ or $\phi_3$, the nodes are
along the direction orthogonal to the SDW axis. Note that although
the nodal structure of our SDW gap function resembles that of a
$p$-wave symmetry, the SDW order that we found is completely on-site
and inversion symmetric.

\subsection{No-full-gap `theorem' in the two band model}
In the following we argue that the Dirac nodes are topologically
stable as long as the SDW order satisfies three conditions (a brief
account of this argument was presented in Ref.\cite{FaPRL}): (1)
collinear order(denote the magnetization direction by $\hat n$) (2)
inversion (about the Fe site $\mathcal{I}$) symmetry and (3)
Effective time reversal symmetry $\mathcal{TR}' =\mathcal{SR}(\hat
n\rightarrow -\hat n)\circ\mathcal{TR}$ obtained by combining time
reversal and spin reversal ($\mathcal{TR}$ is time-reversal and
$\mathcal{SR}(\hat n\rightarrow -\hat n)$ is the 180$^{\circ}$ spin
rotation which flips the direction of magnetization). These three
conditions are naturally satisfied by a $(\pi,0)$ collinear SDW which
is consistent with experiments\cite{Neutrons}, and the mean-field
SDW that we find also satisfies these conditions.

Since we focus on the consequences of the nontrivial band structure,
we begin by turning on a very weak SDW order:  $\hat
M=\sum_{\vec k}\sigma^z_{\alpha\beta} M_{ab}(\vec k)d^{\dagger}_{a\alpha,\vec
k}d_{b\beta,\vec k+(\pi,0)}$ (where we have rotated the
magnetization to the $S_z$ direction).  We therefore
need to consider degenerate perturbation after folding the Brillouin
zone by $(\pi,0)$. The band crossing between the two bands defined
in equation \ref{eq:dispersion}, i.e. $E_{-}(\vec k)$ and
$E_{+}(\vec k+(\pi,0))$ forms a loop in momentum space. In the
perfect nesting case these loops are identical to the electron or
hole fermi surfaces. Let us focus on the loop around $(0,0)$, and
denote the momenta on this loop by $\vec{k}^*$: thus $E_{-}(\vec
k^*) =E_{+}(\vec k^*+(\pi,0))$. Let us call the hole pocket
wave-functions on this loop $|\psi_h(\vec k^*\rangle =
|\psi_{-}(\vec k^*)\rangle$ and the electron pocket wave-function
$|\psi_e(\vec k^*\rangle = |\psi_{+}(\vec k^*+(\pi,0))\rangle$. Now
the degeneracy on this loop is lifted when the matrix elements of
the SDW order between these two kinds of states: $m(\vec
k^*)=\langle \psi_h(\vec k^*)|\hat M|\psi_{e}(\vec k^*)\rangle$ is
non zero.

Now, $\mathcal{I}$ symmetry requires $M_{ab}(-\vec k)=M_{ab}(\vec
k)$. Also, given the winding of the hole Fermi surface wavefunction
shown in Figure \ref{fig:perfect-nesting}b, under inversion we have
$|\psi_h(-\vec k^*\rangle = - |\psi_h(\vec k^*)\rangle$ while
$|\psi_e(-\vec k^*\rangle = |\psi_e(\vec k^*)\rangle$. Putting this
together we have $m(-\vec k^*)=-m(\vec k)$. In addition,
$\mathcal{TR}'$ symmetry requires $M_{ab}(\vec k)$ to be real hence
so is $m(\vec k^*)$. We thus conclude that $m(\vec k^*)$ must have
at least two sign changing points, $K$ and $-K$, on the band
crossing loop. These are the two of the Dirac nodes in the SDW.
Similarly there are another two Dirac nodes on the band crossing
loop around $(0,\pi)$.

Generically the nodes are not at the chemical potential, and leads
to Fermi pockets. However, a Fermi pocket deriving from a node is
known, e.g. from the context of graphene, to be different from a
regular Fermi pocket. In particular, electrons acquire a nontrivial
Berry's phase of $\pi$ on circling such a nodal pocket. We have thus
argued for the stability of the Dirac nodes in SDW based on the
two-band model Eq.(\ref{eq:2-band}). We now show that a similar
result holds for the more realistic five band model, again the
topology of the quadratic bands that touch at the $\Gamma$ point are
responsible for this result.

\section{Nodal SDW in the 5-band model}
While the two band model serves as a useful guide to the
nontrivial physics in the SDW state, it differs from LDA
calculations of the electronic structure of these
materials\cite{LDA} in important ways. This is easiest to see in the
unfolded
 Brillouin zone scheme, with a single Fe atom per unit cell (here
 unit translations along the $x$ and $y$ axes are followed by
 reflections in the $xy$ plane\cite{LeeWen}). While the two band
 model has a hole pocket at $\Gamma=(0,0)$ and another one at $(\pi,\pi)$,
 the LDA calculation predicts two hole pockets around the $\Gamma$ point.
 The electron pockets although centered around the same locations in both
 cases,  acquire a $d_{xz}+d_{xy}$ character at $(0,\pi)$ and
a $d_{yz}+d_{xy}$ character at $(\pi,0)$ in the LDA calculations.
Hence they are also rather different from the two band model that
does not include the $d_{xy}$ orbitals. A 5-band hopping Hamiltonian
including all the iron d-orbitals is required to capture the fermi
 surface topology of the LDA calculation.

In this section we will show that even in the five band model, the
SDW is necessarily gapless (nodal-SDW), despite these important
differences. An important role here is played by the fact that the
two hole pockets at the $\Gamma$ point are derived from the
$d_{xz},d_{yz}$, which are precisely the orbitals that enter the two
band model.  In this case we will prove that there must be at least
two Dirac nodes close to fermi level in the SDW phase.

\subsection{The mean-field study of SDW in the 5-band model}
We again apply the trial density matrix method to study the
$(\pi,0)$ SDW instabilities of the 5-band model. We take the 5-band
hopping Hamiltonian $H_0$ from  Kuroki{\it et al.}\cite{LDA} (Eq.(1)
and Table I). We then turn on a on-site interaction in the following
form\cite{LeeWen}:
\begin{align}
 H_{I}&=U\sum_{i,a}n_{ia\uparrow}n_{ia\downarrow}+(U-2J)\sum_{i,a< b} n_{ia}n_{ib}\notag\\
&+J\sum_{i,a< b}^{\alpha,\beta=\uparrow,\downarrow} d_{ia,\alpha}^{\dagger}d_{ib,\beta}^{\dagger}d_{ia,\beta}d_{ib,\alpha}\notag\\
&+J\sum_{i,a<
b}\big(d_{ia,\uparrow}^{\dagger}d_{ia,\downarrow}^{\dagger}d_{ib,\downarrow}d_{ib,\uparrow}+h.c.\big)\label{eq:5-band-model}
\end{align}
{where $a,b=1,2\ldots 5$ label the 5 orbitals $\{d_{3Z^2-R^2} , d_{XZ} , d_{YZ} , d_{X^2-Y^2},d_{XY}\}$.} We have
parameterized the inter-orbital Coulomb interaction by $U-J$. While
this is strictly expected to hold within the $t_{2g}$ and $e_g$
levels \cite{Castellani}, to reduce the number of interaction
parameters, we assume it for the five band model as well.

\begin{figure}
 \includegraphics[width=0.2\textwidth]{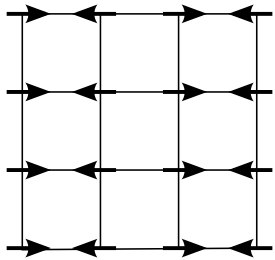}
\caption{The real space pattern of the $(\pi,0)$ SDW on the Fe-atom square lattice.}
\label{fig:sdw-pattern}
\end{figure}

We only consider the on-site SDW order, because the interactions
taken to be on-site. Then, the $(\pi,0)$ SDW order parameter with
spins assumed to point along the $z$ axis induces the following mean
field term in the Hamiltonian:
\begin{equation}
H_{SDW}=\sum_{i}e^{i\pi
 i_x}\sum_{a,b=1}^{5}M_{ab}(d_{i,a\sigma}^{\dagger}\sigma^z_{\sigma\sigma'}d_{i,b\sigma'})
\end{equation}
which is parameterized by $M_{ab}$ a Hermitian matrix with 25 real
parameters. The orbital structure can leads to many different SDW
states which break symmetry in different ways. Energetically, we
find that the preferred state always has the same symmetry as the
regular $(\pi,0)$ SDW shown in Fig.\ref{fig:sdw-pattern}. That is, the
state breaks time reversal ($\mathcal{TR}$) and 180$^o$ spin
rotation about an axis perpendicular to the ordering direction
$\mathcal{SR}$, but preserves their combination ($\mathcal{TR'}$).
Also the state is invariant under inversion ($\mathcal{I}$) which is
a 180$^\circ$ rotation in the $x-y$ plane, the $P_x$ reflection around the $x$-axis crossing an Fe atom
(which actually combines with $z\rightarrow -z$ reflection and is a 3-D 180$^\circ$ rotation around the $x$-axis), and similarly the
$P_y$ reflection ($P_x\circ P_y = \mathcal{I}$). At the end of this subsection we comment on other
possible states, that break different symmetries.

We now discuss details of the mean field solution, as the on-site
interaction strength is varied. Since this is not accurately known,
we note the resulting ordered moment and Fermi surface topology in
each case, which may be directly compared with experiments. The
Hund's coupling $J$ is assumed to be about 20\% of $U$.

\begin{figure}
 \includegraphics[width=0.28\textwidth]{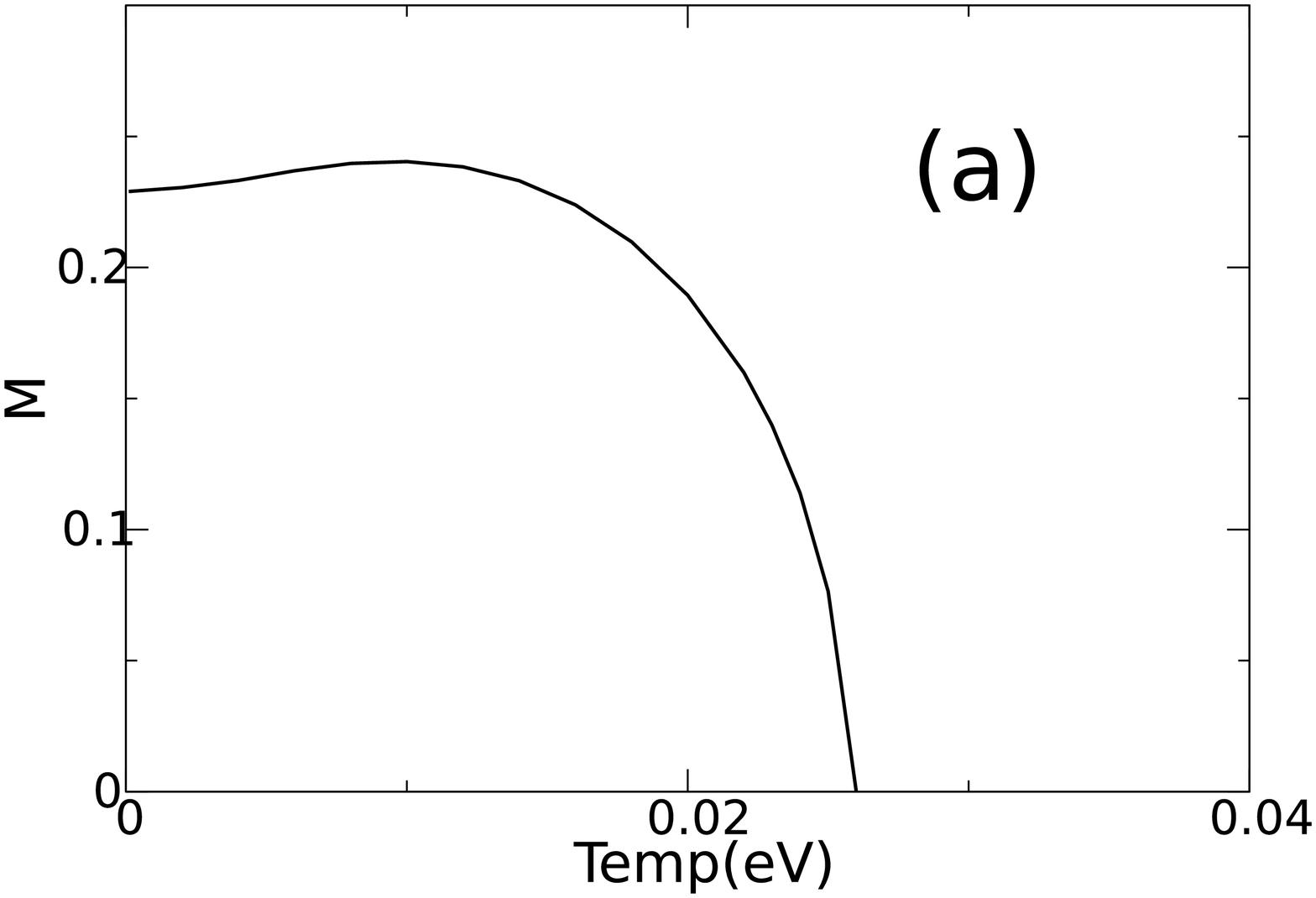}\;\;\;\;\includegraphics[width=0.15\textwidth]{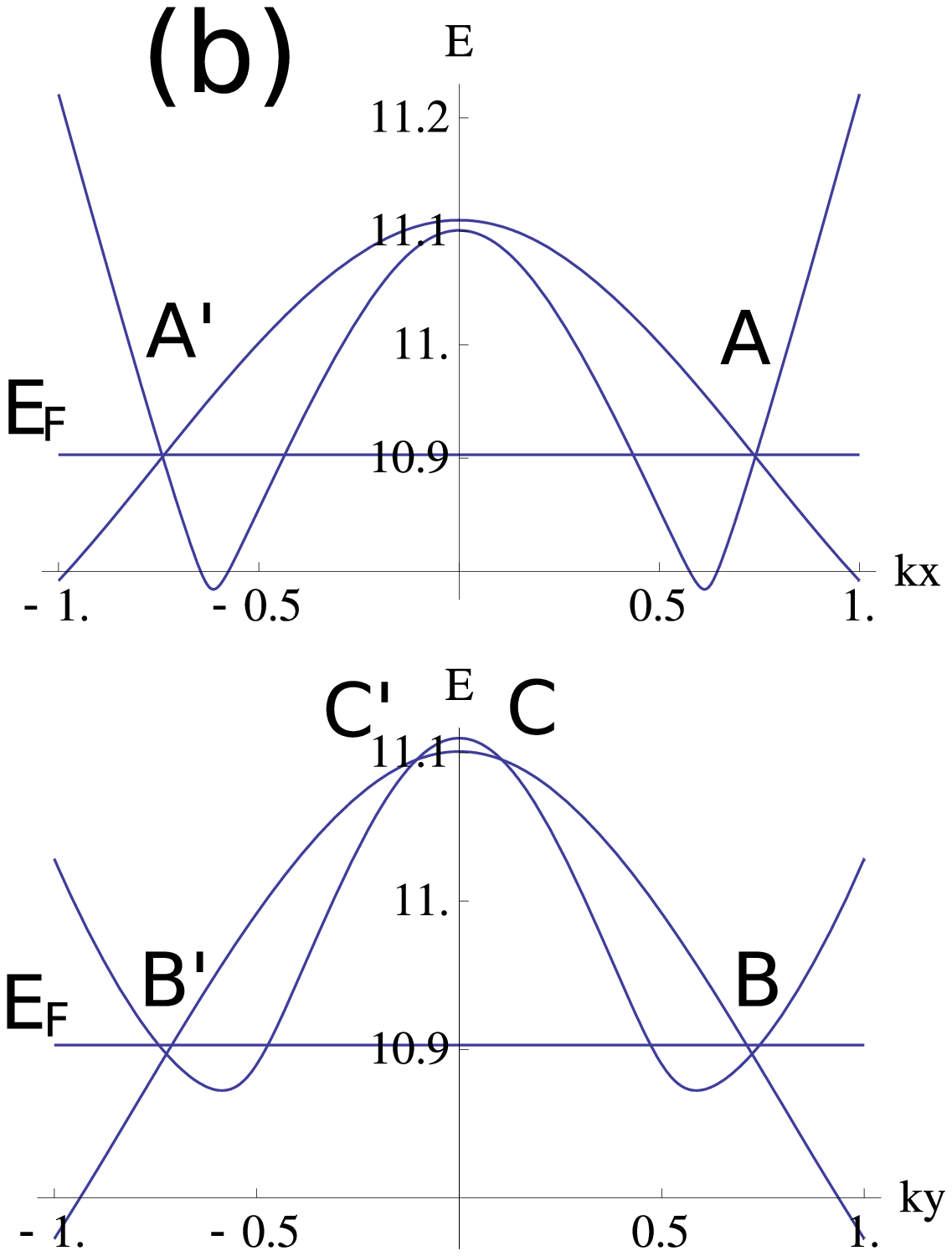}\\\smallskip
 \includegraphics[width=0.15\textwidth]{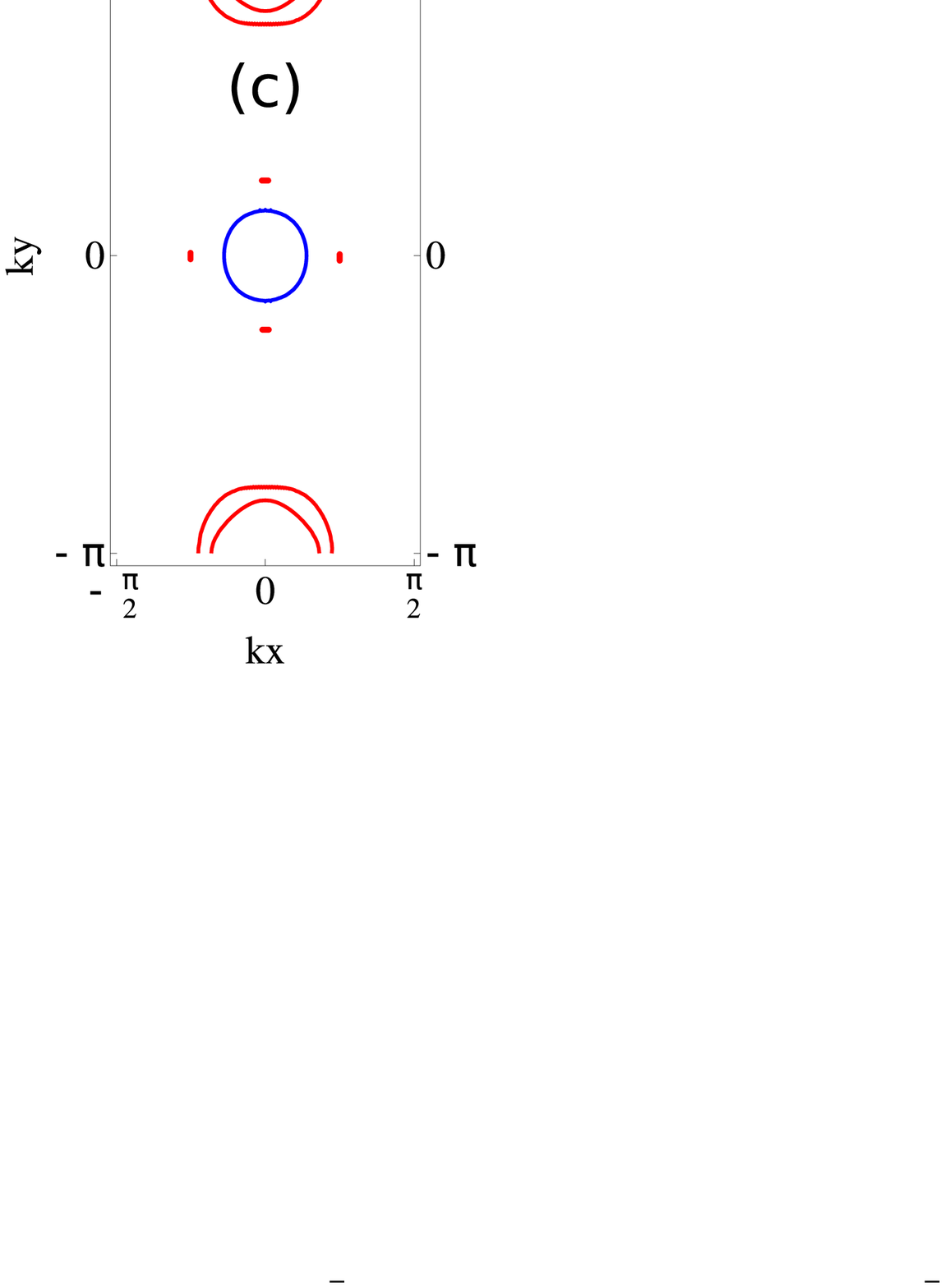}\;\;\includegraphics[width=0.15\textwidth]{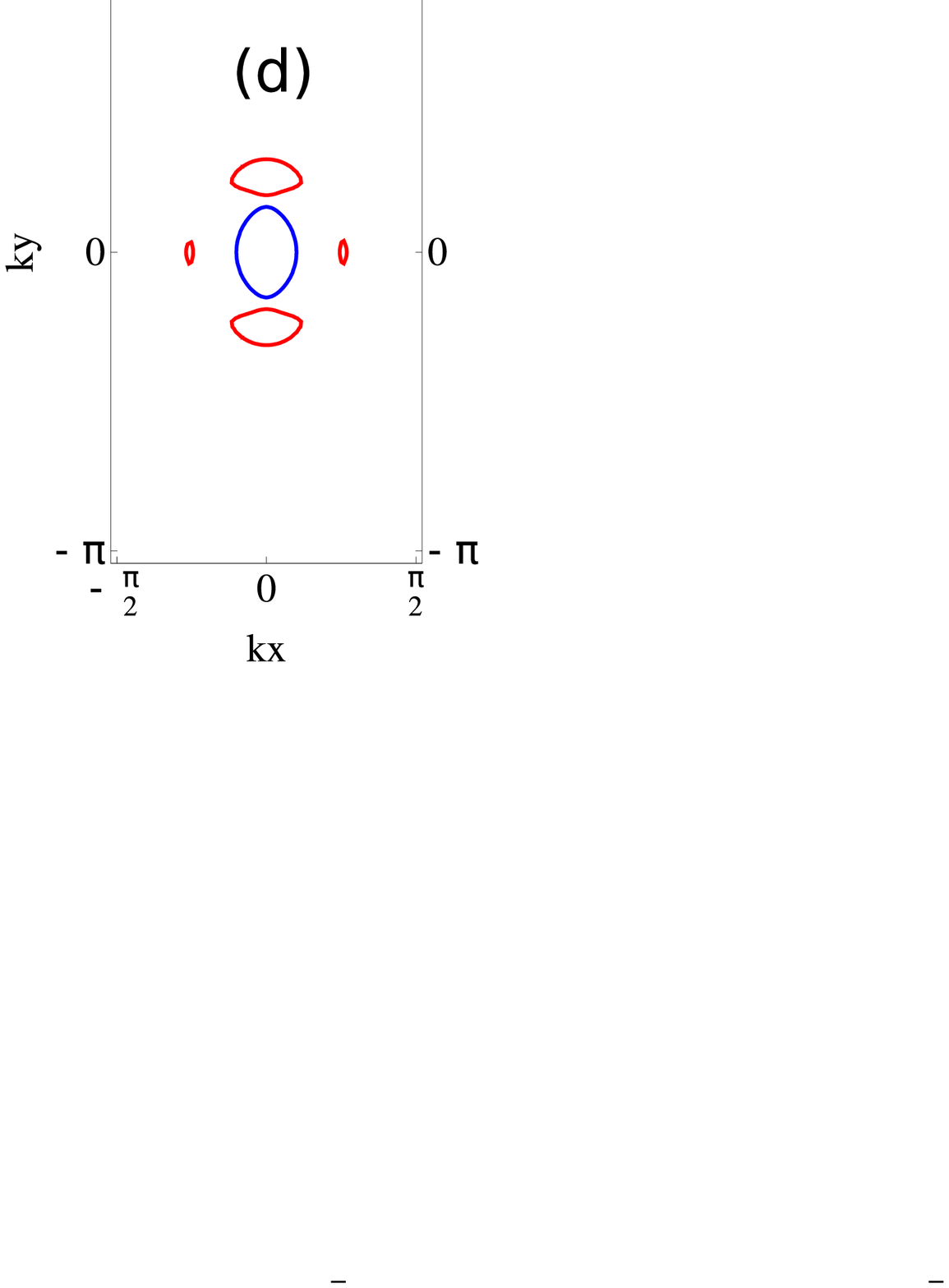}\;\;\includegraphics[width=0.15\textwidth]{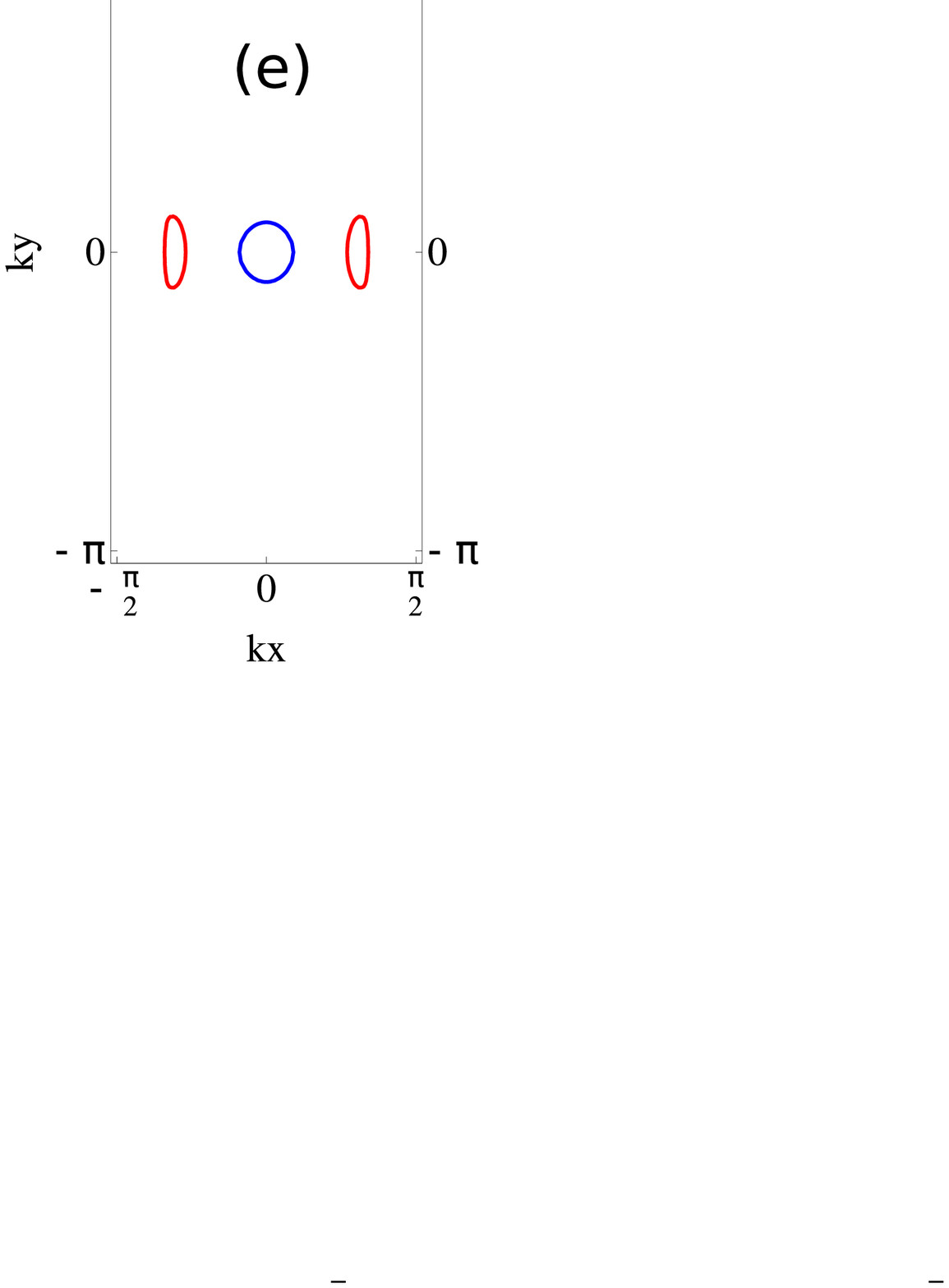}
\caption{(color online) (a) Magnetization with $U=1$eV and $J=0.2$eV,  as a
function of temperature (computation is performed on a 40 by 40
lattice with periodic boundary.) (b) T=0 dispersion in this SDW
state showing the two low energy bands along $k_x$ and $k_y$ axes
around $(0,0)$. There are four nodes close to fermi energy: $A$ and
$A'$ on the $k_x$ axis and $B$ and $B'$ on the $k_y$ axis. The
double degenerate point at $(0,0)$ of the 5-band hopping Hamiltonian
is split into two Dirac nodes, $C$ and $C'$, on the $k_y$ axis. We
also plot the zero temperature fermi surfaces (blue pockets are
hole-like and red pockets are electron-like) with (c) $U=1$eV and
$J=0.2$eV  and with (d) $U=1.2$eV and $J=0.25$eV  and with (e) $U=1.4$eV and $J=0.3$eV.
} \label{fig:5-band-magnetization}
\end{figure}

(1) U=1.0eV, J=0.2eV. The temperature evolution of the net SDW
magnetization $M=\langle\sum_a
d_{ia,\uparrow}^{\dagger}d_{ia,\uparrow}-d_{ia,\downarrow}^{\dagger}d_{ia,\downarrow}\rangle$
(assuming $g=2$ this is in unit of Bohr magneton) from mean field
theory is shown in Fig.\ref{fig:5-band-magnetization}(a). The mean
field transition temperature obtained is $T_c=0.026$eV while the
zero temperature magnetic moment obtained is $\sim 0.23\mu_B$. The
latter is consistent with neutron scattering experiments on
$LaOFeAs$ \cite{Neutrons}. In Fig.\ref{fig:5-band-magnetization}(c)
we present the fermi surfaces of the zero temperature half-filled
SDW phase with these parameters. Note that the double degenerate
point at $(0,0)$ of the 5-band hopping Hamiltonian is split into two
Dirac nodes, $C$ and $C'$, on the $k_y$ axis. The electron pocket at
$(0,\pi)$ still intersect with fermi level and contributes
significantly to the density of states. However, on increasing the
interaction strength this feature is suppressed.  Stronger repulsive
interactions will tend to gap out more of the fermi surface and the
electron pockets at $(0,\pi)$ can be fully gapped out. We note that
the area occupied by electron pocket is $3.5$\% of the magnetic
Brillouin zone (defined by $-\pi/2<k_x<\pi/2$ and $-\pi<k_y<\pi$).
The hole pockets, of course, occupy the same area.

(2) $U=1.2$eV and $J=0.25$eV. In this case we obtain a somewhat
larger low temperature moment $\sim 1.04\mu_B$.  In
Fig.\ref{fig:5-band-magnetization}(d) we plot the fermi surfaces of
the zero temperature SDW. As compared to the previous cases, we see
that the fermi surfaces around $(0,\pi)$ disappears and the fermi
level adjusts itself to form one hole pocket around $(0,0)$ and four
electron pockets. The hole pocket occupies $2.2$\% of the magnetic
Brillouin zone while the area of the electron pockets on the $k_x$
axis is $0.1$\% each, and of those on the $k_y$ axis is $1.0$\%
each.

The electron pockets on $k_x$ axis and on $k_y$ axis are
fundamentally different in that the $k_x$-pockets arise from the
Dirac nodes $A$ and $A'$ below them, and thus protected. On the
other hand the $k_y$-pockets can be easily gapped out by turning on
interactions because they are simple band bottoms (after $B$($B'$) and C($C'$) annihilate each other, which already happens in the current case).

(4) $U=1.4$eV and $J=0.3$eV. The zero temperature moment is now
large $\sim 2.3\mu_B$. Now, the two electron pockets along the $k_y$
axis are completely gapped out and we only have a small hole pocket
around $(0,0)$, and two electron pockets on the $k_x$ axis nearby
(see Fig.\ref{fig:5-band-magnetization}(e)). The fermi surface
topology and large moment obtained in this case are the closest to
the LDA results\cite{LDA}. The hole pockets occupy $1.4$\% of the
magnetic Brillouin zone, while the electron pockets occupy $0.7$\%
each.

 Note, for the larger interaction strengths, the area occupied
by the residual Fermi surface in the SDW state is very small,
typically a few percent. In general, interactions that drive the SDW
formation would tend to lower this area. One may naturally expect
that this would rapidly lead to a fully gapped state on increasing
$U$. However, as explained in detail in the next section, there is
an intrinsic mechanism that blocks such a fully gapped state. A
combination of symmetry and band topology necessarily leads to a
gapless SDW state, over a wide range of coupling strengths. This
provides a 'natural' protection of the small pockets that appear
here and in LDA calculations, which have now been observed in
magnetic oscillation experiments \cite{Lonzarich}.

{\em Other Possible Orders:} Before we conclude the mean field study
of SDW orders, we comment on other kinds of $(0,\pi)$ orders that
could be stabilized with onsite interactions. The 25 parameter
$M_{ab}$ matrix admits a plethora of different orders, which may be
separated into the following four classes, according to the
symmetries of the resulting SDW Hamiltonian $H_{SDW}$:
\begin{itemize}
 \item[(i)] $\mathcal{TR'}$ even, $\mathcal{I}$ even and $P_x,P_y$ even. (6 parameters)
 This is the SDW state that was considered above. The no-full-gap theorem discussed below, applies to this case.
\item[(ii)] $\mathcal{TR'}$ even, $\mathcal{I}$ even and $P_x,P_y$ odd. (3 parameters)
\item[(iii)] $\mathcal{TR'}$ even and $\mathcal{I}$ odd. (6 parameters)
\item[(iv)] $\mathcal{TR'}$ odd. (10 parameters) This case is rather exotic because $\mathcal{TR}$ is even but $\mathcal{SR}$ is odd. This is similar to the symmetry of a spin-hall
insulator\cite{kane:226801}.
\end{itemize}

To compare the relative stabilities of these different states, we
choose $U=1$eV and $J=0.2$eV and first perform an unbiased
minimization of all the 25 parameters. We find that class (i) is
always the low free energy solution and with the highest
$T_c=0.026$eV. Even if we suppress order parameter (i) by hand, we
find the system has no instability towards (ii), (iii) and (iv) down
to 0.0001eV. (Computations were performed on a 40 by 40 lattice with
periodic boundary.) We conclude that SDW (i) is the low free energy
phase in the model Eq.(\ref{eq:5-band-model}) and is consistent with
the ordered pattern observed in experiments. Hence, we do not pursue
studying these other kinds of SDW order.

\subsection{No-full-gap 'theorem' in the 5-band model}
In this section we explain why the nodal SDW found in the mean field
study appears. We assume that the SDW has the symmetries in (i)
above, as found in mean field theory. Briefly, we use reflection
symmetry along the $k_x=0$ and $k_y=0$ to label bands with a
reflection eigenvalue. Bands connected to the electron pocket and a
hole pocket are forced to have opposite eigenvalues and hence do not
split in the SDW state along these lines, leading to a gapless
state. An important role is played by the band touching at the
$k=(0,0)$ point, as in the two band model. While this reasoning
holds for weak SDWs, we extend it to include strong SDW
instabilities, where the location of the band intersections can
migrate to the $\Gamma$ point. Even in this case so we show that at
least two gapless Dirac nodes will remain.

\begin{figure}
 \includegraphics[width=0.35\textwidth]{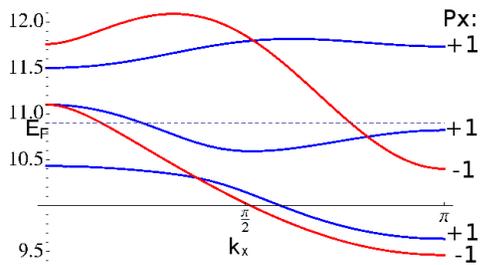}
\caption{(color online) We present the $P_x$ eigenvalues (blue for $+1$ and red for
$-1$) of the 5 bands along $k_x$ axis.}
\label{fig:px}
\end{figure}

{\it Weak SDW Limit:} The reflections $P_x$ and $P_y$ around the $x$ and $y$ axes passing through the Fe atoms (which actually combine with $z\rightarrow -z$ reflection and are 3-D 180$^\circ$ rotations),
act on the orbital basis of
$\{d_{3Z^2-R^2},d_{XZ},d_{YZ},d_{X^2-Y^2},d_{XY}\}$ as follows (note that our definition of $X$ and $Y$ axes given in Fig.\ref{fig:hoppings} is different from the definition in Kuroki, {\it et al.}\cite{LDA} by a 90$^\circ$ rotation.),
\begin{align}
 T_{P_x}:&\begin{pmatrix}1&0&0&0&0\\0&0&1&0&0\\0&1&0&0&0\\0&0&0&-1&0\\0&0&0&0&1\end{pmatrix},& T_{P_y}:&\begin{pmatrix}1&0&0&0&0\\0&0&-1&0&0\\0&-1&0&0&0\\0&0&0&-1&0\\0&0&0&0&1\end{pmatrix}
\end{align}
Spin is left invariant as we ignore spin-orbit interactions. We
first consider the $P_x$ reflection symmetry. Along $k_x$ axis the
Bloch Hamiltonian is $P_x$ symmetric and the wave-function should be
eigenstates of $P_x$ with eigenvalues $\pm 1$. We thus can simply
present the eigenvalue of $P_x$ of each band as shown in
Fig.\ref{fig:px}. If we focus on the three bands close to fermi
level, we find that the electron pocket is $P_x$ odd, the large hole
pocket is $P_x$ even and the small hole pocket is $P_x$ odd. Since
the relevant SDW orders are $P_x$ even,
the SDW induced gap along $k_x$ axis between the electron pocket and
the large hole pocket must vanish, i.e. there is a band crossing.
These are labeled $A$ and $A'$ in the example of
Fig.\ref{fig:5-band-magnetization}b. We note that the bands
corresponding to the large hole pocket and the small hole pocket
must have opposite $P_x$ eigenvalues and as result the nodes must
exist no matter whether the electron pocket is $P_x$ even or odd.
The simplest way to understand this is  to note the double
degenerate wavefunctions at $(0,0)$ are nothing but $d_{xz}$ and
$d_{yz}$. Hence, these band touchings play a crucial role here, as
in the two band model.

Similarly there also must be nodes along the $k_y$ direction by
studying the $P_y$ eigenvalues along the $k_y$ axis. We find that
the electron pocket is $P_y$ odd, while the large hole pocket is
$P_y$ even and small hole pocket is $P_y$ odd. Therefore we expect
the SDW gap vanishes along $k_y$ direction between the electron
pocket and the large hole pocket as well. These are labeled $B$ and
$B'$ in the example of Fig.\ref{fig:5-band-magnetization}b.
Hence, in the limit of a weak SDW, the Fermi energy will inevitably
cross at least one of the protected bands, and a full gap cannot
result. Note, the existence of pockets is independent of how good
the nesting is, in contrast with conventional SDWs which are gapless
only due to the absence of nesting. A further distinction is that
some of the gapless pockets in our case result from Dirac nodes.

{\em Strong SDW:} The above four band crossings are obtained
assuming the SDW is a weak perturbation. If SDW is strong can these
nodes annihilate and lead to a fully gapped SDW?  Note, in the
presence of $\mathcal{TR'}$ and $\mathcal{I}$ symmetry, a band
crossing can only be removed via annihilation with a partner. Since
the band crossings appear close to the $\Gamma$ point, it is
particularly important to address whether they can annihilate by
coming together at that point. For completeness, it is important to
note that in addition to the four band crossings discussed above,
there are two additional ones that arise when $C_{4v}$ is broken in
the SDW state. The double vortex at $(0,0)$ is split into two Dirac
nodes with the same chirality when the SDW order is turned on.
Therefore there are totally 6 Band crossings around $(0,0)$ in the
weak SDW limit (labeled as A,B,C and A',B',C' in the mean field
example of Fig.\ref{fig:5-band-magnetization}b.). In the Appendix we
prove an important theorem based on a topological argument ensuring
these 6 nodes can never annihilate in pairs at $(0,0)$. We show that
this is ensured by the fact that the two wavefunctions at $(0,0)$,
which corresponding to $d_{XZ},d_{YZ}$, are both odd under inversion
($\mathcal{I}$). The nodes can only annihilate in sets of four. We
conclude that there must be at least two Dirac nodes left. This
result is topologically stable. Indeed, on going to stronger
interaction strengths this is the state realized in our mean field
study e.g. in Fig.\ref{fig:5-band-magnetization}f.  In that case the
two left-over Dirac nodes are along the $k_x$ axis.

Now we compare the nodal SDW in 2-band model and 5-band model.
One main difference is that the fermi surface topologies and shapes
are quite different. In particular in the 5-band model there may not
be a large anisotropy of conductivity as in the 2-band model because
at least there is a rather circular hole pocket around $(0,0)$. The
positions of Dirac nodes and fermi surface topologies, which may be
easily measured by single crystal ARPES, can serve as a way to
directly detect the validity of 2-band model or 5-band model.

\section{Conclusions}
In this paper we studied the spin density wave (SDW) ground state of
the undoped FeAs compound. We find that the combination of physical
symmetry and the topology of the band structure, naturally
stabilizes a gapless SDW ground state with Dirac nodes. We first
study a popular two-band model due to its simplicity, where this
mechanism is manifest. We also study the more realistic five-band
model, where the same result obtains. These two rather different
models share a key topological feature of the band structure: the
double degeneracy at $\vec{k}=(0,0)$ enforces a wavefunction winding
around this point in the Brillouin zone.

In both models we perform the mean-field study allowing for all
possible collinear magnetic orders at $(\pi,0)$, and  find the inversion and reflection parities of the
lowest energy magnetic ordered phase. We then show that the SDW ground
state in both models have stable Dirac nodes protected by the
inversion symmetry and the topology of the band structure. These
Dirac nodes are close to Fermi level and thus may be directly
observable in ARPES experiments and might also control the low
energy thermodynamic and transport properties of compound. They
arise due to the vanishing of the SDW matrix elements along a high
symmetry line in the Brillouin zone, which leaves the Fermi surfaces
ungapped in this direction. We also proved a general result on the
stability of Dirac nodes against pairwise annihilation in an
inversion symmetric system (Appendix \ref{app}) which may be applied
to more general situations. While strong interactions tend to
increase the SDW gap and reduce the Fermi pocket area, the nodal
nature of the SDW does not allow a full gap to open over a wide
range of interaction strengths. Hence, one expects to be left with
small residual Fermi surface pockets, which naturally explains the
small Fermi surface areas (0.52\% and 1.38\% of the Brillouin zone with two Fe atoms per unit cell) observed in magnetic oscillation
experiments \cite{Lonzarich}.

Although we find stable Dirac nodes in the SDW ground state in both
the two and five band models, the number and the locations of the
fermi pockets are different. These differences can serve as ways to
determine which is a better model of the material. Effective low
energy theories of the FeAs materials should ideally incorporate the
nodal nature of the SDW state, which may also have important
consequences for other phases in this system.

We acknowledge useful discussions with Cenke Xu. A.V. would like to
thank Leon Balents for stimulating conversations, Steve Kivelson for
pointing out Ref.\cite{Kamihara} at an early stage and support from
LBNL DOE-504108, NSF-DMR 0645691. D. H. L. was supported by DOE
DE-AC02-05CH11231.

\appendix
\section{Annihilation Condition for Dirac nodes}\label{app}

We consider the general question of when a pair of Dirac nodes can
come together and annihilate, to give rise to a non-singular band
structure. We assume the existence of both {\em time reversal} and
{\em inversion} symmetry. We consider the case where a pair of Dirac
nodes come together at a point $\mathcal M$ in momentum space -
which is constrained by the symmetries above to be invariant under
inversion. When the Dirac nodes are near this point, we can restrict
attention to just the two bands that make up these nodes. At the
$\mathcal M$ point, they can be labeled by their eigenvalues under
inversion $I_1=\pm 1$ and $I_2=\pm 1$. Below we show that only if
$I_1I_2= -1$ can the pair of nodes annihilate. Otherwise, they
necessarily lead to a band touching with quadratic dispersion at the
$\mathcal M$ point, when they are brought together.

To derive this result, we use a Berry's phase formula to fix the
inversion eigenvalue of the $\mathcal M$ point states before and
after node annihilation. This places the required constrain on the
inversion eigenvalues, if the nodes are to annihilate. We first
define the inversion parity $\sigma(C)$ of a half loop $C$
connecting two points $P$ and $P'$, which are mapped to one another
by inversion. We start with the wavefunction of one of the bands at
$P$, $|\Psi_P\rangle$, which can be taken to be real given that we
have both time reversal and inversion symmetry.  This is evolved
adiabatically along the contour $C$ to give the real wavefunction
$|\Psi_P'\rangle$ at point $P'$. Clearly this is an eigenstate of
the Bloch Hamiltonian at this point in the Brillouin zone. A
separate way to obtain the eigenstate at $P'$ is to apply the
inversion operation on the state at $P$: $I|\Psi_P\rangle$. Again,
the inversion operation can be constructed to yield a real
wavefunction. Hence, these two wavefunctions can at most differ by a
sign,
\begin{equation}
|\Psi_P'\rangle =\sigma(C)I|\Psi_P\rangle
\end{equation}
which is the inversion parity $\sigma(C)$ of the curve $C$. Note
that since $I^2=\mathbf{1}$ $\sigma(C)$ is independent of the
direction of $C$. Although it depends on the band index, the band
label is suppressed for clarity.
\begin{figure}
 \includegraphics[width=0.2\textwidth]{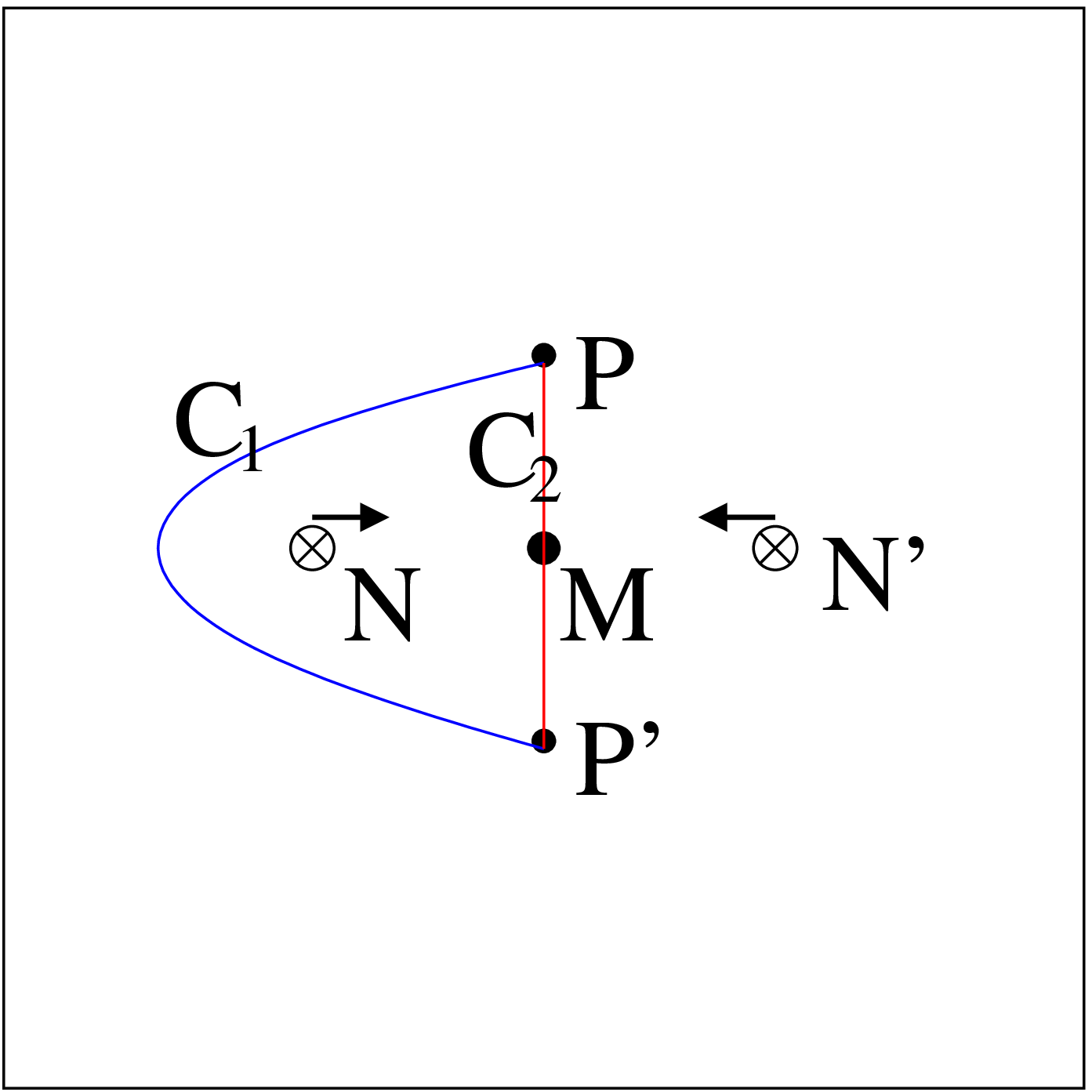}\;\;\includegraphics[width=0.2\textwidth]{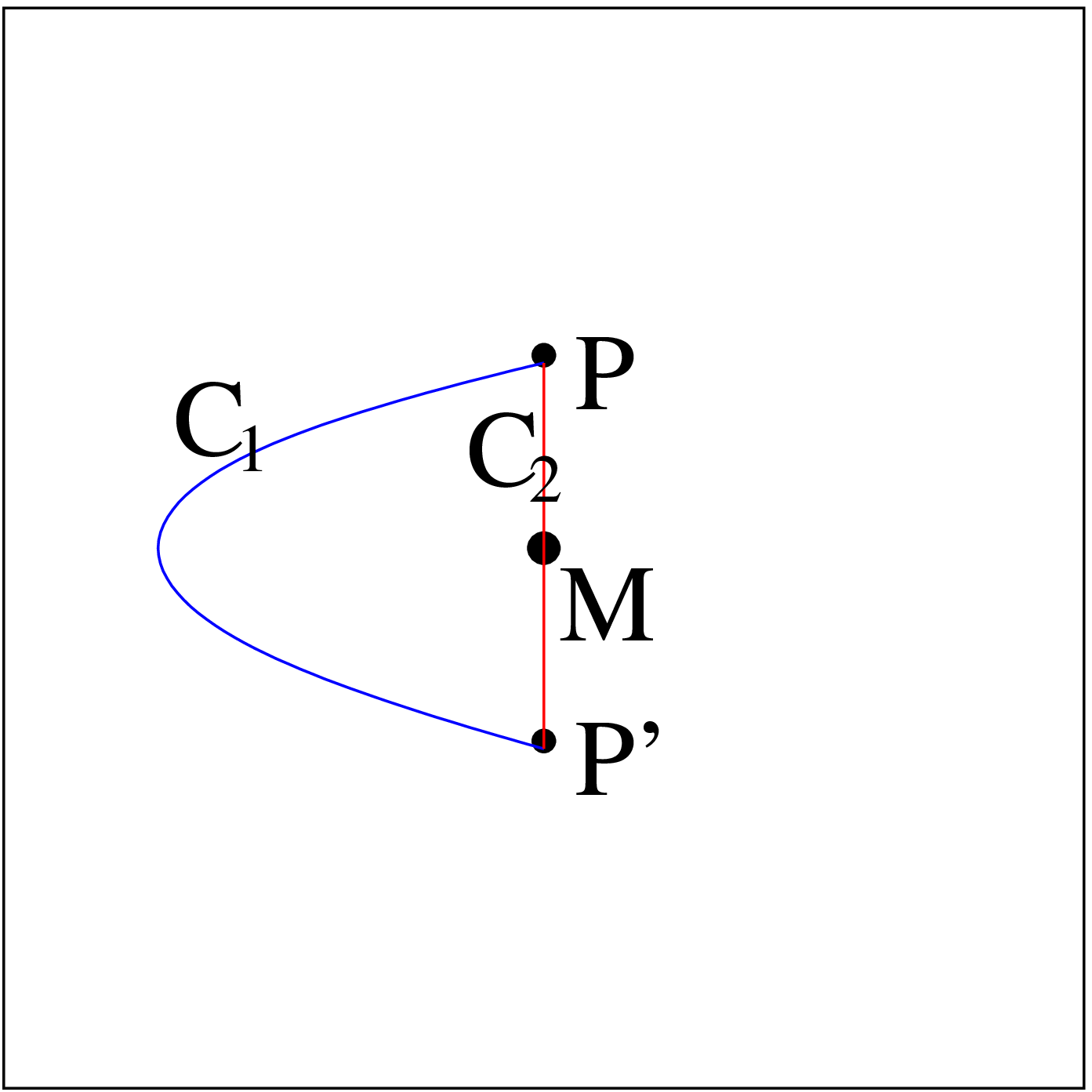}
\caption{We show the paths $C_1$ and $C_2$ connecting two points $P$
and $P'$ which is inversion image of each other: (a) before
annihilation and (b) after annihilation. $C_2$ crosses the inversion
symmetric point $M$, and before annihilation $C_1+C_2$ encloses one
single Dirac node $N$.} \label{fig:annihilation}
\end{figure}

Now consider two nodes $N$ and $N'$ being brought together at $
\mathcal M$ and choose two points $P$ and $P'$ along the
perpendicular direction of this path. As shown in
Fig.\ref{fig:annihilation} we choose two paths $C_1$ and $C_2$
connecting $P$ and $P'$. Initially, $C_1+C_2$ encloses a single
Dirac node at $N$, and as a result we must have
$\sigma(C_1)\sigma(C_2) = -1$ because the wavefunction must wind by
$\pi$ around a Dirac point. Now {\em assume} that the nodes
annihilate on being brought together. Now, $C_1+C_2$ encloses no
singularity, so, the inversion parities $\sigma'(C)$ after this
operation satisfy $\sigma'(C_1)\sigma'(C_2)=+1$. However, since the
wavefunctions along $C_1$ evolve smoothly during the annihilation we
have: $\sigma'(C_1)=\sigma(C_1)$. Therefore we  must have
$\sigma'(C_2)=-\sigma(C_2)$. Note, by shrinking  the curve $C_2$ we
can approach the $\mathcal M$ point. Then, the inversion parity of
the curve simply becomes the eigenvalue under inversion of the
wavefunction at the $\mathcal M$ point: ($I_1,\,I_2$). Therefore we
conclude that in the node annihilation process, the inversion
eigenvalue of each of the two states at the $\mathcal M$ point
changes sign. This is only possible if they have opposite signs to
begin with, $I_1I_2=-1$. In that case they can simply pass through
each other, and the net result will be a sign change of the
inversion eigenvalue of the higher and lower energy states. However,
if they both have the same sign, $I_1I_2=+1$, it is not possible to
affect a sign change. In this case, our assumption that the nodes
annihilate is invalid - in fact a pair of bands with quadratic
dispersion will touch at the $\mathcal M$ point.

 If
the two bands have opposite inversion eigenvalues, then the
inversion matrix in the two bands is $\tau^3$. And the inversion
symmetric real Hamiltonian around $M$ must be able to expand as
$(a\delta k_x)\tau^1+(\epsilon +b \delta k_x^2+c \delta
k_y^2+d\delta k_x \delta k_y)\tau^3$ to the quadratic order after
choosing the $k_x$ axis to be along direction connecting the two
nodes. We immediately see that depending on the sign of $\epsilon$,
the Hamiltonian either has two or zero band touching nodes. This
indicates that if the two bands have opposite inversion eigenvalues,
the two nodes can always annihilate at $\mathcal M$.

Finally we note that the SDW state in the five band model meets the
conditions required for the above analysis to hold.  Inversion
$\mathcal{I}$ is a symmetry of the system, and the role of time
reversal is played by $\mathcal{TR'}=\mathcal{SR}(\hat n\rightarrow
-\hat n)\circ \mathcal{TR}$(defined in text) symmetric system, where
$\hat n$ is the direction of collinear SDW.
Let us choose the orbital basis $d_a$ $(a=1\ldots n)$ to be
eigenfunctions of inversion and label the eigenvalues to be $I_a$.
We then define the $\mathcal{I}\circ\mathcal{TR}$ even basis $\tilde
d_a$ in the following fashion: if $I_a=1$ then $\tilde d_a=d_a$, and
if $I_a=-1$ then $\tilde d_a=id_a$. For a collinear SDW with
$\mathcal{TR'}$ symmetry the Hamiltonian in $\tilde d_a$ is purely
real, and so are the eigenfunction in the momentum space. In this
basis, the arguments presented above can be made, leading to the
conclusion that a pair of Dirac nodes cannot be annihilated at the
$\Gamma$ point. Hence, since we begin with six nodes in all, there
will always be a leftover pair that is stable.



\end{document}